\documentstyle[12pt,preprint]{aastex}
\begin{document}

\title{New Luminous ON Spectra from the Galactic O-Star Spectroscopic Survey}

\author{Nolan R.\ Walborn}
\affil{Space Telescope Science Institute, 3700 San Martin Drive, Baltimore, MD 21218}
\authoremail{walborn@stsci.edu}

\author{Nidia I.\ Morrell}
\affil{Las Campanas Observatory, Carnegie Observatories, Casilla 601, La Serena, Chile}
\authoremail{nmorrell@lco.cl}

\author{Rodolfo H.\ Barb\'a\altaffilmark{1}}
\affil{Departamento de F\'{\i}sica y Astronom\'{\i}a, Universidad de La Serena, Cisternas
1200 Norte, La Serena, Chile}
\authoremail{rbarba@dfuls.cl}

\author{Alfredo Sota}
\affil{Instituto de Astrof\'{\i}sica de Andaluc\'{\i}a--CSIC, Glorieta de la
Astronom\'{\i}a s/n, 18008 Granada, Spain}
\authoremail{sota@iaa.es}

\altaffiltext{1}{Visiting Astronomer, Las Campanas Observatory.}

\newpage

\begin{abstract}
Two new ON supergiant spectra (bringing the total known to seven) and one new ONn giant (total of this class now eight) are presented; they have been discovered by the Galactic O-Star Spectroscopic Survey.  These rare objects represent extremes in the mixing of CNO-cycled material to the surfaces of evolved, late-O stars, by uncertain mechanisms in the first category but likely by rotation in the second.  The two supergiants are at the hot edge of the class, which is a selection effect from the behavior of defining N~\textsc{iii} and C~\textsc{iii} absorption blends, related to the tendency toward emission (Of effect) in the former.  An additional N/C criterion first proposed by Bisiacchi et~al.\ is discussed as a means to alleviate that effect, and it is relevant to the two new objects.  The entire ON supergiant class is discussed; they display a fascinating diversity of detail undoubtedly related to the complexities of their extended atmospheres and winds that are sensitive to small differences in physical parameters, as well as to binary effects in some cases. Serendipitously, we have found significant variability in the spectrum of a little-known hypergiant with normal N, C spectra selected as a comparison for the anomalous objects.  In contrast to the supergiants, the ONn spectra are virtual (nitrogen)-carbon copies of one another except for the degrees of line broadening, which emphasizes their probable unique origin and hence amenability to definitive astrophysical interpretation.   
\end{abstract}


\section{Introduction}

The visibility of material processed by their own interior nuclear reactions at the surfaces of stars is a remarkable diagnostic that will contribute to unique definitions of their evolutionary status and histories, once the complex mechanisms involved can be identified and understood.  An inverse N/C,O dichotomy in certain OB spectra was announced and discussed by Walborn (1971, 1976).  Unfortunately, reliable quantitative analysis of these spectra has lagged those results by decades, because of the complexity of especially the supergiant atmospheres and winds; only recently have thorough analyses of substantial samples begun to appear (e.g., Rivero Gonz\'alez et~al.\ 2011, 2012a,b; Bouret et~al.\ 2012; Martins et~al.\ 2015a,b).  There is a growing consensus that the rare extreme cases of the phenomenon bracket systematic trends of nitrogen enrichment among massive stars, such that, e.g., the majority of \textit{morphologically} normal OB supergiants are \textit{physically} enhanced, while the OBC and OBN spectra correspond to opposite extremes, as originally suggested by Walborn (1976).  Of course, that can be established only quantitatively, not only because of the range of degrees, but also because morphological detection of even extreme cases depends on the availability of suitable criteria, which is nonuniform and produces selection effects.        

The Galactic O-Star Spectroscopic Survey (GOSSS; Ma{\'\i}z Apell\'aniz et~al.\ 2011) ambitiously proposes to observe the blue-violet spectra of every accessible O~star in the Galaxy, i.e., a few thousand, with homogeneous resolving power ($\sim$2500) and signal-to-noise ($\sim$200).  Nearly 450 verified or new classifications have already been published (Sota et~al.\ 2011, 2014; Papers~I \& II, respectively) and about as many more are in progress.  Such a project is guaranteed to reveal new categories of spectra as well as individual objects of special interest, and that has already occurred (Walborn et~al.\ 2010, 2011).  The first of those references addresses curious differences in the relative intensities of N~\textsc{iii} and C~\textsc{iii} emission features among Of spectra, which may or may not be related to chemical abundances since selective excitation effects are involved.  The second expands upon the ONn category of nitrogen-enhanced, rapidly rotating, late-O giant spectra, further increased by one in this paper.  Two new ON~supergiants are likewise significant because of their rarity and the need to disentangle possibly multiple physical mechanisms within the category.  Thus, these three new objects are presented and discussed in their respective contexts here, along with some related developments in their classification criteria.  The new objects will be included in the next installment of the GOSSS classification lists (Ma{\'\i}z Apell\'aniz et~al.\ 2016, in press; Paper~III).
     
\section{Observations}

Spectroscopic observations of all stars but two discussed here have been obtained with the Boller \& Chivens spectrograph attached to the 2.5m du~Pont telescope at the Las Campanas Observatory (LCO) in Chile. The Marconi No.~1 $2048\times515$ $13.5\mu$-pixel detector was in use with that instrument.  We used a 1200~lines~mm$^{-1}$ grating centered at 4700~\AA\ and had the slit width set to 150~microns, corresponding to 1\farcs26 on the sky and 1.75 pixels on the detector.  This instrumental configuration produces a resolution of $\sim$1.7~\AA\ as measured from the FWHM of the comparison lines.  The typical peak signal-to-noise ratio (S/N) per 2-pixel resolution element ranges from 150 to 200, with just a few cases significantly below or above.  Dome flats and bias frames were obtained during the afternoon prior to each observing night as well as twilight flats at sunset, and He-Ne-Ar comparison-lamp exposures were recorded before or after each target was observed.  All of these data pertain to the Galactic O-Star Spectroscopic Survey (GOSSS; Ma{\'\i}z Apell\'aniz et~al.\ 2011).  They have been adjusted to the uniform survey resolving power of 2500 and are incorporated into the GOSSS database.  

HD~191781 and BD~+36$^{\circ}$~4063 in the northern hemisphere were observed with the Albireo spectrograph at the 1.5m~reflector of the Observatorio de Sierra Nevada (OSN), Granada, Spain.  The 1800~lines~mm$^{-1}$ grating provided a spectral scale of 0.62~\AA\ per pixel.  Further details of these data are given by Sota et~al.\ (2011, 2014). 

\section{Results}

\subsection{The ON Supergiants}

\begin{deluxetable}{lcllrrccc}
\rotate
\tabletypesize{\small}
\tablecolumns{8}
\tablewidth{0pc}
\tablecaption{Observational and Measured Data for the Supergiants}
\tablehead{
\colhead{Name}&\colhead{RA}&\colhead{DEC}&\colhead{SpT}&\colhead{$V^{\rm a}$}&
\colhead{$B-V$}&\colhead{EW($\lambda\lambda$4511--15)$^{\rm b}$}&\colhead{EW($\lambda$4650)$^{\rm b,c}$}&\colhead{EW Ratio}}
\startdata
HDE 323110& 17:21:15.79 &  $-$37:59:09.6&ON9~Ia&9.67&+1.00&$0.44\pm0.01$&$0.53\pm0.01$&$0.83\pm0.02$\\
HDE 328209& 16:29:19.16& $-$44:28:14.2&ON9~I&9.77&+0.80&$0.74\pm0.01$&$0.56\pm0.01$&$1.32\pm0.02$\\
HD 123008 & 14:07:30.65 & $-$64:28:08.8 & ON9.2~Iab&8.84 &+0.37&$0.62\pm0.01$&$0.32\pm0.01$&$1.94\pm0.01$\\
HDE 269896& 05:37:49.11 & $-$68:55:01.7&ON9.7~Ia$^+$&11.36&0.00 &$0.37\pm0.01$&$0.29\pm0.01$&$1.28\pm0.02$\\
HD 105056& 12:05:49.88&$-$69:34:23.0&ON9.7~Iae&7.34&$-$0.14 &$0.50\pm0.01$&$0.36\pm0.01$&$1.39\pm0.02$\\
BD+36$^\circ$4063 &20:25:40.61&+37:22:27.1&ON9.7~Ib&9.69&+0.98&$0.40\pm0.02$&$0.17\pm0.01$&$2.35\pm0.03$\\
HD 191781 & 20:09:50.58 & +45:24:10.4   & ON9.7~Ia &9.53 & +0.64&$0.44\pm0.01$&$0.35\pm0.01$&$1.26\pm0.02$\\
HD 173010&18:43:29.71&$-$09:19:12.6&O9.7~Ia$^+$&9.22&+0.62&$0.51\pm0.01$&$1.27\pm0.01$&$0.40\pm0.02$\\
HD 173010&            &           &B0~Ia$^+$&       &    &$0.54\pm0.01$&$1.45\pm0.01$&$0.37\pm0.02$\\
HD 104565 & 12:02:27.79 & $-$58:14:34.4 & OC9.7~Iab&9.26 & +0.36&$0.24\pm0.01$&$1.62\pm0.01$&$0.15\pm0.02$\\
\enddata
\tablenotetext{a}{Multiple photometric sources can be found in SIMBAD, whence the values listed here have been selected.  The $B$~magnitude given for HDE~269896 at the top of its SIMBAD page is in error.}  
\tablenotetext{b}{Equivalent widths and their errors are given in \AA.}
\tablenotetext{c}{Our measurements of ``$\lambda$4650'' necessarily include Si~\textsc{iv}~$\lambda$4654, because at our resolution the latter is incipiently resolved from the C~\textsc{iii} blend only in optimum cases of line quality and weak C~\textsc{iii}.}
\end{deluxetable}

\begin{figure}
\epsscale{1.0}
\plotone{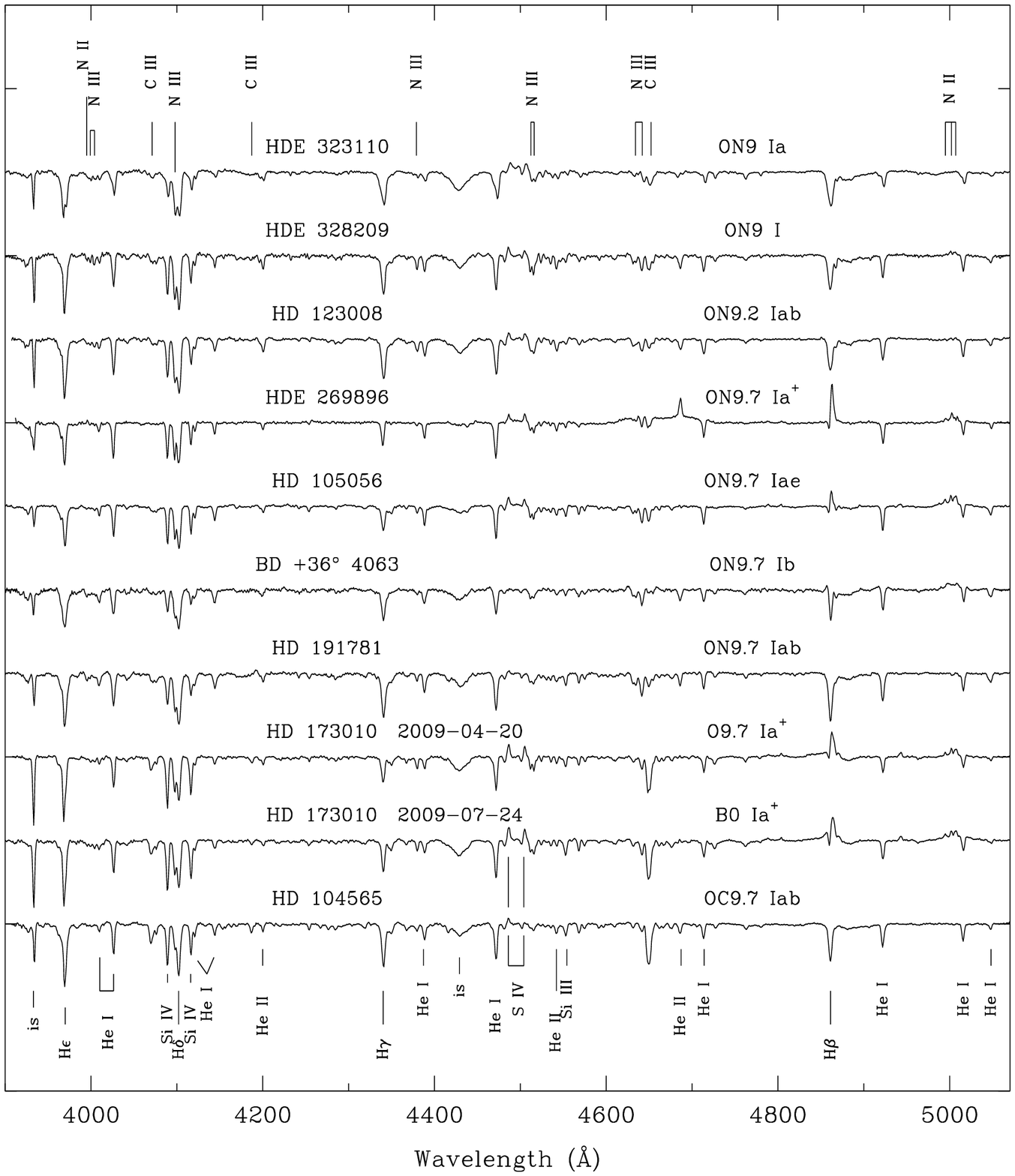}
\caption{The spectra of the known ON supergiants in order of spectral type advancing from top to bottom.  Below are plotted examples of morphologicially normal and OC spectra for comparison.  See Walborn \& Howarth (2000, Figs.~3--4) for the wavelengths of the identified spectral features.}
\end{figure}

The blue-violet spectra of the seven ON supergiants are displayed in order of advancing spectral type in Figure~1, together with morphologically normal and OC supergiants below for comparison.  Data for these objects are listed in Table~1, in the same order as the figure.

\subsubsection{The Spectra of HDE 323110 and 328209}

The two new members have somewhat earlier types (higher temperatures), which at first glance had excluded them from the category.  That is because the original definition of an ON~star required the N~\textsc{iii}~$\lambda\lambda$4634--4640--4642 absorption blend (hereafter ``$\lambda$4640'') to be stronger than C~\textsc{iii}~$\lambda\lambda$4647--4650--4652 (``$\lambda$4650''), which is not the case in these spectra.  However, the other N~III features are exceedingly strong, including the $\lambda\lambda$4511--4515 blend discussed later as an additional criterion for the class, and especially $\lambda$4097 in the shortward wing of H$\delta$.  C~\textsc{iii} is also significantly weaker than normal.  The relative weakness of $\lambda$4640 is a result of filling by the Of selective emission, which increases toward earlier types (and higher luminosities), resulting in a strong selection of the ON classification toward the latest O~types.  We have three observations of HDE~328209, from 2011 March~24, 2013 May~30, and 2014 April~26, among which all spectral features are essentially identical; the second of these is shown in Fig.~1. 

The exact spectral types of these two stars are not well defined due to somewhat conflicting criteria.  Both the He~\textsc{ii}~$\lambda$4200/He~\textsc{i}~$\lambda$4144 and He~\textsc{ii}~$\lambda$4541/He~\textsc{i}~$\lambda$4387 ratios should have values of unity at type O9, increasing toward earlier and decreasing toward later types.  However, in both of these spectra the first of those ratios is slightly larger (tending toward type O8.5) while the latter is slightly smaller (type O9.2; Sota et~al.\ 2014).  Thus, the compromise typeO9 is adopted for both.  In the first ratio, the discrepancy is most likely caused by blending with enhanced N~\textsc{iii}~$\lambda\lambda$4196, 4200.  In addition, HDE~323110 is an SB2, with blended, shortward secondary components visible at the He~\textsc{i} lines in Fig.~1.

Unfortunately, the luminosity classes of the two spectra are likewise problematic.\break HDE~323110 appears to have a very weak P~Cyg profile at the principal criterion He~\textsc{ii}~$\lambda$4686, which is consistent with class~Ia.  (An alternative interpretation could be a blend of lines from the two SB components, weaker in the longward spectrum which would still indicate Ia.) In contrast, in HDE~328209, the $\lambda$4686 absorption (which weakens with increasing luminosity also due to the Of effect) is only slightly weaker than He~\textsc{i}~$\lambda$4713, which calls for class~Ib, in substantial disagreement with the exceedingly strong Si~\textsc{iv}~$\lambda\lambda$4089, 4116 absorptions that are appropriate for Ia.  Hence, only a relatively indeterminate luminosity class of I can be assigned to this spectrum.    

A further interesting feature of these spectra is the prominent quartet of weak, similarly spaced absorption lines near $\lambda$4000.  Its bookends are the usual N~\textsc{ii}~$\lambda$3995 and He~\textsc{i}~$\lambda$4009, which are at the low-ionization extreme of these spectra, while the two intermediate lines, favored by the ionization and nitrogen enhancement, are N~\textsc{iii}~$\lambda\lambda$3999, 4004 that have similar intensities to the outer two for those reasons.  This configuration can be recognized in the other ON~I spectra briefly discussed next, albeit with differing relative intensities corresponding to their respective spectral types.

There is also interesting prior and current external information about both of these stars.  The anomalous N/C spectra of HDE~323110 (= LSS~4103) were accurately first reported by Vijapurkar \& Drilling (1993); our later recognition of this characteristic is independent and thus confirmatory of theirs.  Moreover, in agreement with the above description of line profiles in our spectrogram, the OWN Survey (Barb\'a et~al.\ 2010) has discovered that this object is a short-period, eclipsing, interacting SB2, which will be presented in the context of that program.  HDE~328209 was reported to be a runaway star by Moffat et~al.\ (1998) on the basis of HIPPARCOS data.    

\subsubsection{Other ON~I Stars}

All of the known ON supergiants are included in Fig.~1 for comparison with the two new ones; some of their characteristics are briefly described here.  While on the one hand they constitute a homogeneous spectroscopic category, on the other when examined closely they show considerable diversity of detail, and some of them may have diverse physical natures.  They are ordered by the  He~\textsc{ii}~$\lambda$4541/Si~\textsc{iii}~$\lambda$4552 classification ratio, the unit value of which defines spectral type~O9.7, although some range is allowed.  

HD~123008 is the earliest/hottest of these spectra, both by that criterion and by the two He~\textsc{ii}/He~\textsc{i} horizontal classification ratios discussed above in the context of the new objects.  Actually, those ratios display the same small discrepancy shown by the new members, likely for the same reason, and arguably this spectrum could be classified the same, but it will not be revised here.  Martins et~al.\ (2015b) derive log(N/H)~+~12~=~9.1 in this spectrum, and N/C~=~61 by number.  The corresponding values in the Sun are 7.9 and 0.25, respectively, as cited by Evans et~al.\ (2004).  Some other quantitative results as available will be noted below for comparison.    

HDE~269896 is a hypergiant in the Large Magellanic Cloud ($M_V -8.1$) and displays several remarkable features related to its extreme luminosity that are unique among this set.  Most notable is the strong He~\textsc{ii}~$\lambda$4686 emission line that is very rare at such a late O~type.  The range of ionizations in emission is also noteworthy, as discussed by Corti et~al.\ (2009), from N~\textsc{ii} through H$\beta$ to the He~II line, probably indicating a very extended atmosphere.  It is interesting to note in the present context that $\lambda$4640 is not stronger than $\lambda$4650 in this spectrum, either, but that is due to the high luminosity rather than to a higher temperature as in the new members.
Evans et~al.\ (2004) quote log(N/H)~+~12~=~8.3 and N/C~=~7.9 for this object, but Corti et~al.\ (2009) found that increasing the former value to 8.9 together with other small parameter changes provides a far better fit to the observed spectrum, including emission vs.\ absorption features.  Sanduleak (Sk) $-66^\circ$~169 also in the LMC has a very similar O9.7~Ia$^+$ spectrum except with morphologically normal CNO spectra; for it Evans et~al.\ (2004) derive corresponding values of 7.95 and 4.5, while quoting LMC H~\textsc{ii} region values of 7.1 and 0.13, respectively, for comparison. 

HD~105056 and BD~+36$^\circ$~4063 may be the jokers in this deck.  The former is possibly a low-gravity subdwarf, because of its large distance from the Galactic plane, which would be even more extreme if it had a Population~I supergiant luminosity (Walborn et~al.\ 1980).  The latter is a known SB undergoing active mass transfer (Williams et al. 2009) and its spectrum is significantly variable; HDE~323110 may be a similar system.  The unit He~\textsc{ii}~$\lambda$4686/He~\textsc{i}~$\lambda$4713 ratio in our observation of BD~+36$^\circ$~4063 corresponds to luminosity class~II in normal spectra.

HD~191781 lies at the extreme late/cool edge of the O9.7 type, almost but not quite B0.  The spectral analysis of Martins et~al.\ (2015b) yields log(N/H)~+~12~=~8.9 and N/C~$>$~15 in this object.   

\subsubsection{Morphologically Normal and OC Supergiants}

HD~173010 is a highly luminous object; the intensities of the 
Si~\textsc{iv} absorption lines are unprecedented, leading to the Ia$^+$ classification despite the lack of He~\textsc{ii}~$\lambda$4686 emission as in HDE~269896.  Nevertheless, its C, N spectra are morphologically normal, so it is displayed here as a comparison for the anomalous objects.  The N~\textsc{iii} lines are very strong, except for $\lambda$4640 which must be filled in.  But C~\textsc{iii} is also extremely strong, cf.\ $\lambda$4650.  Perhaps this star had a lower initial rotational velocity than those above?  The long-unidentified S~\textsc{iv} emission lines at $\lambda\lambda$4485, 4503 (Werner \& Rauch 2001) are also extremely strong.  One wonders if the metal abundances might be supersolar in this object toward the inner Galaxy.  

Surprisingly, we have found that the spectral type and (not independently) the Si ionization of HD~173010 are variable.  Two observations are shown in Fig.~1, from 2009 April~20 (upper) and 2009 July 24 (lower).  From the He~\textsc{ii}~$\lambda$4541/Si~\textsc{iii}~$\lambda$4552 classification ratio, the first spectral type is O9.7, similarly to the adjacent HD~191781, but the second is substantially later and must be classified as B0 (cf.\ the atlas of standards in Sota et~al.\ 2011).  Concurrently, the Si~\textsc{iv}/Si~\textsc{iii} ionization ratio changes in the same sense, an effect we have not seen previously.  Note also the asymmetrical broad emission wings at H$\beta$, as discussed and explained in several 
B~hypergiants and Luminous Blue Variables by Walborn et~al.\ (2015); this profile also appears somewhat variable.  Further temporally resolved investigation of this object is underway at high spectral resolution in the OWN survey.  Clearly, we still have much to learn about massive stellar atmospheres and evolution.      

The OC supergiant HD~104565 is the antithesis of the ON: all the C~\textsc{iii} features are exceedingly strong, while N~\textsc{iii} is vanishingly weak for the spectral type.  The deficiency of $\lambda$4097 between Si~\textsc{iv}~$\lambda$4089 and H$\delta$ is especially striking and is the hallmark of the class; even in normal spectra it is comparable to the 
Si~\textsc{iv} line.  If the initial rotational velocity is a prime cause of mixing, then it must have been extremely low in this star.  Martins et~al.\ (2016, in preparation) derive log(N/H)~+~12~=~8.3 and N/C~$<$~1.0 in this spectrum.

For comparison with the earlier spectral types of the two new ON supergiants, we cite Figure~11 of Sota et~al.\ (2011), which presents a complete luminosity sequence of O9~spectra with morphologically normal CNO.  They include $\alpha$~Cam, O9~Ia, for which Martins et~al.\ (2015a) derive log(N/H)~+~12~=~8.4 and N/C~=~5.  Then we also call attention to HD~152249, OC9~Iab, in Figure~6 of Sota et~al.\ (2014), for which Martins et~al.\ (2015a) derive corresponding values of 8.1 and 0.46, respectively, again in perfect agreement with the morphological predictions.

\subsection{A New ONn Giant}

\begin{figure}
\epsscale{0.5}
\plotone{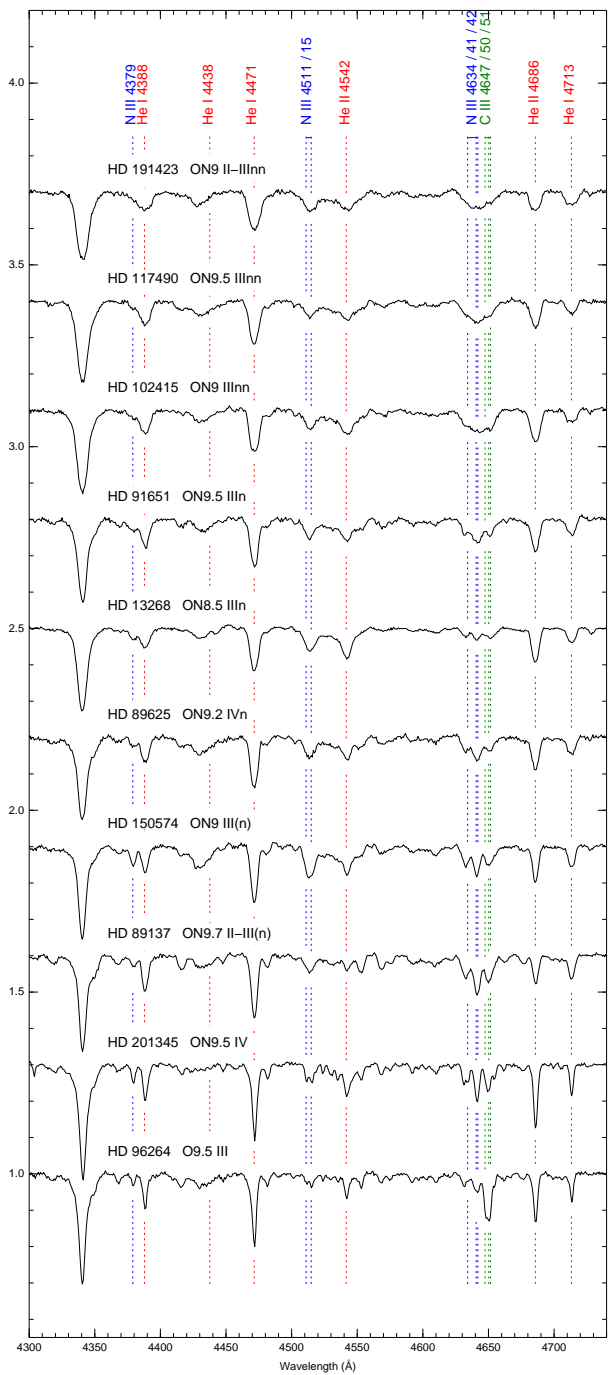}
\caption{The spectra of the known ONn giants in order of increasing $v\sin i$ from bottom to top.  This is Fig.~1 of Walborn et~al.\ (2011) with the new example HD~89625 inserted.  The restricted spectral range is intended to emphasize the behavior of the $\lambda\lambda$4640--4650 blends.  Below are the original ON star HD~201345 (Walborn 1970) with narrower lines, and the morphologically normal giant HD~96264 for contrast.  Figure courtesy of Jes\'us Ma\'{\i}z Apell\'aniz.}
\end{figure}

The ONn class of nitrogen-enhanced, rapidly rotating late-O giants was extensively discussed by Walborn et~al.\ (2011), following its augmentation from the GOSSS data.  It is remarkably homogeneous, comprising a very small range of spectral types among which the principal variable is the high projected rotational velocity, which in combination with the nitrogen enhancement has striking progressive morphological effects on the $\lambda\lambda$4640--4650 blends.  The purpose of this section is simply to add a further member to the class, HD~89625, ON9.2~IVn, discovered subsequently by GOSSS.  Here it has been incorporated into Figure~2, which is the updated Figure~1 of Walborn et~al.\ (2011).  Its location corresponds to its relative $v\sin i$ in ascending order, between HD~150574 ($\sim$200~km~s$^{-1}$) and HD~13268/91651 ($\sim$300~km~s$^{-1}$).  This seamless interpolation again emphasizes the internal morphological consistency of the class, which likely corresponds to a uniform physical origin.  Unfortunately, there are no quantitative spectroscopic data available for the little known HD~89625,\footnote{This circumstance is likely related to the misclassification of B3/B4~(V) from the Michigan HD catalogue quoted (with a slight variation) by SIMBAD.  A hypothesis for this error can be advanced: the weak, broadened He~\textsc{ii} lines were missed on the photographic objective-prism plate, while the difference between the He~\textsc{i} spectrum and its maximum at type B2 was applied in the wrong direction. As reported in VizieR, B.~Skiff (Catalogue of Stellar Spectral Classifications) has found that the original HD classification of B was upgraded to a relatively accurate B0 on the Extension chart.} which is not even included in the OWN survey yet.  For comparison with the values in Table~2 of Walborn et~al.\ (2011), its Galactic latitude is $-$2.4$^{\circ}$.

The mechanisms producing different degrees of processed material at the surfaces of OB stars are a subject of intense current debate.  Most of the ONn spectra have recently been analyzed quantitatively and further discussed by Martins et~al.\ (2015b).  Perhaps the most vexing variable is the unobservable \textit{initial} rotational velocity, which can be inferred only uncertainly and even somewhat circularly from evolutionary models.  In the ONn class, however, the nitrogen enhancement is very likely related to the \textit{current} rotational velocity.  As discussed by Walborn et~al.\ (2011) and Martins et~al.\ (2015b), a high fraction of the class are spectroscopic binaries, and some are runaways, suggesting a mechanism involving binary evolution.  This is an obvious problem ripe for intensive theoretical investigation with the hope of a well defined solution.   

\section{Discussion}

\subsection{Some Statistics}

To date (January 2016) GOSSS has observed a total of nearly 900 O~stars, for 590 of which reviewed or new spectral classifications have been or are about to be published (Sota et~al.\ 2011, 2014; Ma\'{\i}z Apell\'aniz et~al.\ 2016, in press). The former total includes 56 with spectral types in the ranges O9--O9.7~Ib-Ia, of which 7 are ON and 4 OC; and 74 in the ranges O8.5--O9.7~IV--II(n)--nn, of which 8 are ON and none OC.  Thus, the ON supergiants comprise 12.5\% of the total in their spectral-type ranges, and the ONn giants 11\% of their ranges.  Of course, the ``normal'' samples exhibit differing degrees of mixed processed material, including nitrogen-strong (Nstr) and nitrogen-weak (Nwk) spectra that are less extreme than the ON/OC,\pagebreak\ and in particular a majority with some degree of mixing among the supergiants (see Walborn 1976).  The extreme ON/OC spectra are small tails of the overall distributions.

\subsection{A New Classification Criterion for Somewhat Hotter ON~I Spectra}

As discussed early in the history of the OBN/OBC classification dichotomy (Walborn 1976), the recognition of such morphological anomalies is nonuniform in the HRD, depending upon the available features and their normal two-dimensional (temperature, luminosity) behavior.  The $\lambda\lambda$4640--4650 absorption blends in late-O supergiants provide a prime example.  In normal spectra, the longward C~\textsc{iii} blend is much stronger than the adjacent, shortward N~\textsc{iii}.  When the latter becomes the stronger in the rare ON spectra, a striking anomaly results (Fig.~1).  When the two are about equal, the ``Nstr'' notation is used.  However, this criterion functions optimally only in the restricted O9.2--B0.5 range.  At later types, first the N~\textsc{iii} and then the C~\textsc{iii} become too weak and are dominated by O~\textsc{ii} blends at the same wavelengths.  Toward earlier types, the issue is the progressive filling in of the N~\textsc{iii} absorption by the Of emission effect, such that it never exceeds the C~\textsc{iii} even for comparable or greater N enhancements.  Thus, a strong selection effect in the ON classification arises.  It is interesting to note that it also applies to the even rarer case of luminosity exceeding Ia, as in HDE~269896 with He~\textsc{ii}~$\lambda$4686 in emission despite the late-O spectral type; that has been compensated in its ON classification even though the N~\textsc{iii} absorption is only comparable to the C~\textsc{iii}.

Here we have introduced that same compensation for somewhat hotter supergiants, namely HDE~323110 and 328209, in which the two blends are again comparable, while other N~\textsc{iii} absorption features are abnormally intense and C~\textsc{iii} is deficient, as already described above.  Actually, this development has recently been anticipated by the ON8~III((f)) classification of VFTS~819 in 30~Doradus (Walborn et~al.\ 2014, Fig.~22) although $\lambda$4640 is nearly neutral.  Note that other N~\textsc{iii} absorption features in its spectrum, including $\lambda\lambda$4511--4515, are stronger than in the somewhat more luminous O8 spectra adjacent to it in the figure.

The possible relevance of $\lambda\lambda$4511--4515 as an additional indicator of N~enhancement was discussed long ago by Bisiacchi et~al.\ (1982).  While their discrimination of the feature's normal two-dimensional behavior was not entirely satisfying, their proposal deserves reconsideration in the present context.  It would indeed be interesting if accurate quantitative EW measurements of this feature might reveal smaller, continuous degrees of N~enhancement relative to the normal trends over a wider range of O~types, as they discussed.  Perhaps its ratio to C~\textsc{iii}~$\lambda$4650 would provide even higher sensitivity.  As a start, we provide such measurements in Table~1 for the spectra in Figure~1.  It is encouraging that the ratios show monotonic discrimination among the ON, normal, and OC spectra in the expected sense: the ON range (which may well be real) runs from 0.83 to 2.35 with a mean of 1.48~$\pm$~0.19~(m.e.), while the normal spectrum yields 0.4 and the OC 0.15.  Those ratios can be derived from the measurements of Bisiacchi et~al.\ for 8~supergiants with the classifications of which as such we agree (in a number of other cases they list as supergiants objects that are giants in our system; note also their typo for HD~202124): $\alpha$~Cam, 0.43; $\zeta$~Ori, 0.15; HD~75222, 0.19; $\mu$~Nor, 0.08; HD~152249, 0.10; HD~202124, 0.41; 19~Cep, 0.22; HD~225146, 0.10.  Of these, $\zeta$~Ori is Nwk and HD~152249 is OC, while we consider the remainder to have morphologically normal CNO.  Thus there is good to excellent agreement with our measurements in most cases, with the exceptions of $\mu$~Nor and HDE~225146, which have values appropriate for OC while their spectral appearance is compatible with normal CNO, albeit perhaps at the low-N side of the range.  Bisiacchi et~al.\ consider the full O-type spectral grid, not just the small range of this paper.  Walborn et~al.\ (2014, Figures~13 and 15) described the O6.5~V((f))z spectra of VFTS~089 and 761 as Nstr based on the unusual strength of $\lambda\lambda$4511--4515 and other N~\textsc{iii} lines for that spectral type.  Of course, these measurements should be undertaken in the extensive, full GOSSS sample to establish the normal trends and individual deviations from them. Such is beyond the scope of this study but shall be pursued in the future.      

\acknowledgments
We acknowledge Jes\'us Ma\'{\i}z Apell\'aniz (Centro de Astrobiolog\'{\i}a, Madrid) for initiating and implementing the massive Galactic O-Star Spectroscopic Survey; and Emilio Alfaro (Instituto de Astrof\'{\i}sica de Andaluc\'{\i}a) for administrative and funding leadership of GOSSS while it was based at IAA.  An interested and knowledgeable referee offered several comments and questions that enhanced the presentation.  STScI is operated by the Association of Universities for Research in Astronomy, Inc., under NASA contract NAS5-26555.  Publication support was provided by the STScI Director's Discretionary Research Fund.  This research benefited from the thorough reference lists in SIMBAD.  NIM thanks the LCO technical staff for their excellent support during her runs.  RHB acknowledges support from FONDECYT (Chile) via Regular Grant No.~1140076.

\end{document}